\documentclass{WileyMSP-template}

\begin{document}

\pagestyle{fancy}

\title{Self-Referenced Terahertz Semiconductor Dual-Comb Sources}

\maketitle


\author{Ziping Li}
\author{Xuhong Ma}
\author{Kang Zhou}
\author{Chenjie Wang}
\author{Xiaoyu Liao}
\author{Wen Guan}
\author{Shumin Wu}
\author{Han Liu}
\author{Zhenzhen Zhang}
\author{J. C. Cao*}
\author{Min Li}
\author{Ming Yan}
\author{Heping Zeng*}
\author{Hua Li*}



\begin{affiliations}
Dr. Z. Li, X. Ma, K. Zhou, C. Wang, X. Liao, W. Guan, S. Wu, H. Liu, Z. Zhang, Prof. J. C. Cao, Prof. H. Li\\
Key Laboratory of Terahertz Solid State Technology\\
Shanghai Institute of Microsystem and Information Technology\\
Chinese Academy of Sciences\\
865 Changning road, Shanghai 200050, China\\
E-mail: hua.li@mail.sim.ac.cn; jccao@mail.sim.ac.cn

X. Ma, K. Zhou, C. Wang, X. Liao, S. Wu, H. Liu, Prof. J. C. Cao, Prof. H. Li\\
Center of Materials Science and Optoelectronics Engineering\\
University of Chinese Academy of Sciences\\
Beijing 100049, China

W. Guan\\
School of Information Science and Technology\\
ShanghaiTech University\\
393 Middle Huaxia Road, Shanghai 201210, China\\

Prof. M. Li\\
School of Optical Electrical and Computer Engineering\\
University of Shanghai for Science and Technology\\
Shanghai 200093, China\\

Prof. M. Yan, Prof. H. Zeng\\
State Key Laboratory of Precision Spectroscopy\\
East China Normal University\\
Shanghai 200241, China\\
E-mail: hpzeng@phy.ecnu.edu.cn

Prof. H. Zeng\\
Chongqing Key Laboratory of Precision Optics\\
Chongqing Institute of East China Normal University\\
Chongqing 401120, China\\
\end{affiliations}



\begin{abstract}

Employing two frequency combs with a slight difference in repetition frequencies, the dual-comb source shows unique advantages in high precision spectroscopy, imaging, ranging, communications, etc. In the terahertz (THz) frequency range, the electrically pumped quantum cascade laser (QCL) offers the possibility of realizing the compact dual-comb source due to its semiconductor-based chip-scale configuration. Although the active stabilization of a THz QCL dual-comb source was demonstrated by phase locking one of the dual-comb lines, the full stabilization of all dual-comb lines is still challenging. Here, we propose a self-reference method to obtain a fully stabilized dual-comb signal on a pure THz QCL platform. Without using any external locking components, we filter out one dual-comb line and beat it with the whole dual-comb signal, which eliminates the common carrier offset frequency noise and reduces the dual-comb repetition frequency noise. It is experimentally demonstrated that the self-reference technique can significantly improve the long-term stability of the dual-comb signal. A record of the ``maxhold" linewidth of 14.8 kHz (60 s time duration) is obtained by implementing the self-reference technique, while without the self-reference the dual-comb lines show a ``maxhold" linewidth of 2 MHz (15 s time duration). The method provides the simplest way to improve the long-term stability of THz QCL dual-comb sources, which can be further adopted for high precision measurements.

\end{abstract}


\section{Introduction}

The dual-comb spectroscopy employing two optical frequency combs with slightly different repetition frequencies has revolutionized the conventional spectroscopic techniques due to its unique features in the high spectral resolution and fast data acquisition\cite{bernhardt2010cavity,villares2014dual,CoddingtonOptica,dutt2018chip,picque2019frequency,2016terahertz}. Utilizing the slight difference in the repetition frequencies, an equivalent sampling of one frequency comb against the other one can be achieved, which can map the optical spectral data to the microwave frequency range. This time-delayed operation without mechanical scanning gives the dual-comb a natural potential for real-time and high precision spectroscopy\cite{ideguchi2014adaptive,millot2016frequency,zhao2016picometer,LiACSPhoton}.

Since the first demonstration of dual-comb sources working in near-infrared and visible wavelengths, the dual-comb operation frequency range has been extended to other frequencies\cite{millot2016frequency,link2017dual,ideguchi2012adaptive,yan2017mid}. In the terahertz (THz) frequency range, the electrically pumped quantum cascade laser (QCL) is an ideal candidate source for dual-comb operation\cite{villares2014dual,kohler2002terahertz,burghoff2014terahertz,hugi2012mid} due to the large output power\cite{2017multiwatt}, broad spectral coverage\cite{2018heterogeneous,2012ground}, and fast self-detection capability\cite{LiACSPhoton,2016chip,2019chip}. The latter results from the ultrafast carrier dynamics behavior in QCLs without a relaxation oscillation, while offering the unique possibility of on-chip dual-comb and compact dual-comb configurations\cite{LiACSPhoton,2019chip,villares2015chip}. Although the frequency comb and dual-comb operation have been observed in free-running THz QCLs\cite{burghoff2014terahertz,zhou2019ridge,villares2016dispersion}, the improved stability or reduced phase noise by using various locking techniques and/or dispersion engineering are still much in demand. All efforts to improve the comb performance are to control two frequencies of a comb laser, i.e., the repetition frequency $f_{\rm{rep}}$ and carrier offset frequency $f_{\rm{ceo}}$. Once the two frequencies are known, the frequency of each comb line can be fully defined by $f_N$=$f_{\rm{ceo}}$+$Nf_{\rm{rep}}$ with $N$ being the $N$-th order of the frequency comb. For the control of $f_{\rm{rep}}$, the microwave injection locking\cite{barbieri200713,2010injection} and passive saturable absorption\cite{li2019graphene} were successfully implemented to achieve broadband and stable combs. The control of $f_{\rm{ceo}}$ is more challenging. The traditional $f$-$2f$ method\cite{holzwarth2000optical} cannot be used to assess $f_{\rm{ceo}}$ of THz QCL combs due to the limited emission bandwidth. Recent endeavors by employing the coherent sampling and phase seeding techniques with an assistance of femto-second lasers have been successfully demonstrated to lock $f_{\rm{ceo}}$ of a single THz QCL comb\cite{barbieri2010phase,oustinov2010phase}.    
  
Regarding the control of THz QCL dual-comb sources, it is more difficult because the full lock of a dual-comb source is associated with four frequencies, i.e, two repetition frequencies and two carrier offset frequencies. Each noise component of a single comb is directly transferred to the dual-comb lines through the multi-heterodyne beating process. Therefore, the simultaneous locking of four frequencies is extremely hard to be implemented experimentally due to the involved complex locking electronics and facilities. Alternatively, instead of controlling $f_{\rm{ceo}}$ and $f_{\rm{rep}}$ of individual comb lasers, it is more executable to stabilize the dual-comb signal as a whole. This means that it is more feasible to lock the dual-comb carrier offset frequency ($f_{\rm{ceo,dc}}$=$f_{\rm{ceo1}}$-$f_{\rm{ceo2}}$) and the dual-comb repetition frequency ($f_{\rm{rep,dc}}$=$f_{\rm{rep1}}$-$f_{\rm{rep2}}$) because the entire dual-comb signal can be precisely obtained through the multiheterodyne beating. Recently, we have achieved the active stabilization of a THz QCL dual-comb source by phase locking one dual-comb line to a microwave local oscillator employing a traditional phase-locked loop\cite{zhao2021active}. Although the stability of the dual-comb lines was improved, we observed that the noise of the dual-comb lines that were located far away from the locked line (carrier) increased with the increase of the frequency difference. Furthermore, the active stabilization of the dual-comb source involves complex locking ``hardwares". Therefore, it is promising to implement a stabilization or locking of a dual-comb source without using any external locking facilities or components. This ``software"-based stabilization can be achieved by using

computational\cite{burghoff2016computational} 
and adaptive approaches\cite{chen2020adaptive}. In ref. \cite{burghoff2016computational}, it has been shown that the computational approach allows for extracting the phase and timing signals of a multiheterodyne spectrum without any extra measurements or optical elements. And the other approach, i.e., adaptive dual-comb, is based on the automatically triggering data acquisition with a filtered dual-comb line as a sampling clock. In the visible and infrared regions, the adaptive sampling method has been maturely developed for Doppler precision spectroscopic measurements\cite{ideguchi2014adaptive,ideguchi2012adaptive}. Actually, one can note that each dual-comb line shares the same dual-comb carrier and its noise doesn't change the intrinsic dual-comb behaviour. If the common carrier noise is cancelled, the stability of each dual-comb line can be significantly improved. This common-mode-noise-free dual-comb was recently demonstrated in a hybrid dual-comb system which consisted of a QCL comb and an optically rectified (OR) comb with a femto-second laser pump\cite{cappelli2019retrieval}. Since the amplitudes and phases of the OR comb can be taken as a reference, the common-mode-noise-free technique reported in ref. \cite{cappelli2019retrieval} allows for retrieving the phases of the sample comb (QCL comb).

In this work, based on a pure two-QCL system, we propose a self-reference method to obtain a fully stabilized THz dual-comb source. Different from the previously reported active stabilization which strongly relies on the complex phase locking electronics, in this work, without using any external locking components, we filter out one dual-comb line and beat it with the whole dual-comb signal, which finally cancels out the carrier-offset frequency of the dual-comb signal, as well as the noise. It is demonstrated that the self-reference technique can significantly improve the long-term stability of the dual-comb signal, i.e., a narrow ``maxhold" linewidth of $\sim$15 kHz for a time duration of 60 s is obtained, which is more than 130 times smaller than the ``maxhold" linewidth measured from the dual-comb source without a self-reference stabilization.

\section{Experimental setup}

\begin{figure}[!t]
 \centering
 \includegraphics[width=0.78\linewidth]{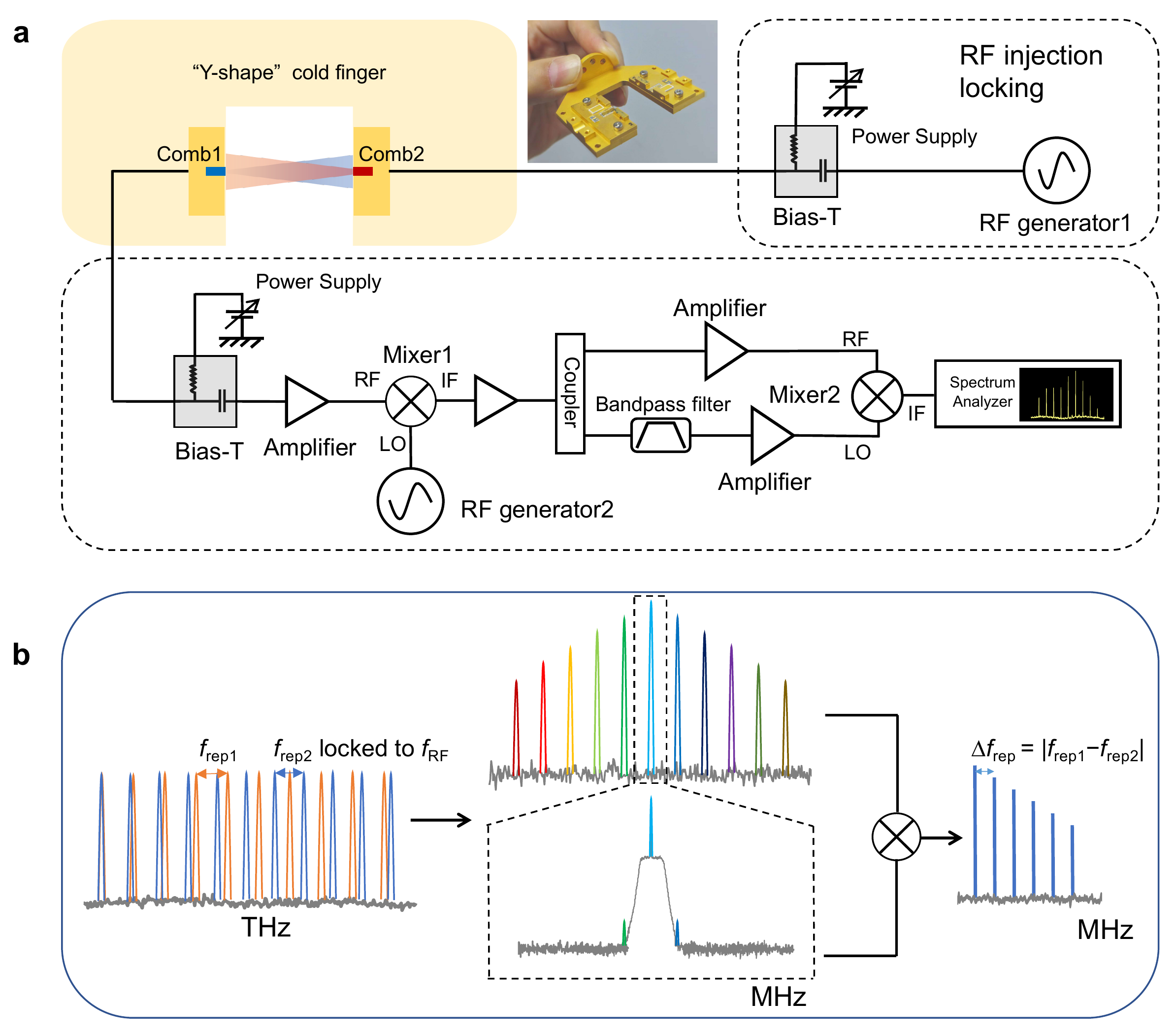}
 \caption{ a) Experimental setup of the self-referenced dual-comb source. The top left, top right, and bottom panels schematically show the THz QCL dual-comb geometry, RF injection locking, and self-reference process, respectively. Comb1 and Comb2 are THz QCLs with a cavity length of 6 mm, which are mounted on a ``Y-shape" cold finger of a continuous-flow helium cryostat. RF generator1 is used to injection lock Comb2 by modulating the laser current at its repetition frequency. RF generator2 acts as a local oscillator (LO) and down-converts the dual-comb signal from the GHz range to MHz range with an assistance of Mixer1. Mixer2 is used to performs self-reference mixing and its intermediate frequency (IF) signal outputs the self-referenced dual-comb. The inset is a photo of the THz QCL dual-comb source mounted on a ``Y-shape" cold finger. b) Frequency synthesis of the proposed self-reference process. $f_{\rm{rep1}}$ and $f_{\rm{rep2}}$ are the repetition frequencies of Comb1 and Comb2, respectively. $f_{\rm{rep2}}$ is locked to $f_{\rm{RF}}$. In the middle panel, the curves with different colors represent the down-covered dual-comb lines in the MHz range and the line selected by a bandpass filter is outlined by a dashed rectangle. The blue lines in the rightmost panel shows the obtained self-referenced dual-comb signal with the common carrier frequency eliminated.}
 \label{setup}
\end{figure}

\textbf{Figure \ref{setup}}a schematically depicts the experimental setup of the proposed self-referenced THz QCL dual-comb. The two THz QCLs (Comb1 and Comb2) with an identical nominal cavity length of 6 mm are mounted on a Y-shape cold finger, same as the geometry reported in ref.\cite{LiACSPhoton}. No optics are used for the optical coupling between the two lasers. A detailed description of the THz QCLs used in this work can be found in the Experimental Section. The top right panel outlined by a dashed rectangle shows the radio frequency (RF) injection locking setup\cite{2010injection,LiOE} which is used to improve the $f_{\rm{rep}}$ stability of Comb2 by modulating its drive current at the repetition frequency of Comb2. Note that here we only implement the injection locking for Comb2 because the laser Comb2 shows much worse $f_{\rm{rep}}$ stability than Comb1 (see figure \ref{comb character}a). To improve the dual-comb signal quality, the RF injection locking is only applied onto Comb2. In principle, if the two laser combs show comparable stability, the injection locking can be removed for simplicity. In this experiment, Comb1 is used as a fast THz detector to measure the multi-heterodyne beating signals employing the laser self-detection mechanism. The multi-heterodyne dual-comb and the inter-mode beatnote signals are extracted from the AC port of the bias-T for the self-reference experiment which is schematically shown in the bottom rectangular box of figure \ref{setup}a. The extracted dual-comb signal in the gigahertz (GHz) frequency range is first amplified and then sent to a mixer to be down-converted to the megahertz (MHz) frequency range. After that, a microwave coupler is used to split the down-converted dual-comb signal into two, one of which is sent to a bandpass filter to filter only one tooth of the lines. Finally, the two signals are amplified again and then sent to a mixer to obtain the self-referenced dual-comb signal which is registered on a spectrum analyser. Note that for the frequency mixing, the filtered signal is acting as a local oscillator (LO).
 
Figure \ref{setup}b shows the frequency synthesis and conversion of the self-referenced dual-comb experiment. The red and blue curves illustrate the emission lines of Comb1 and Comb2 in the THz band, respectively, which shows two different repetition frequencies, i.e., $f_{\rm{rep1}}$ and $f_{\rm{rep2}}$. As mentioned before, $f_{\rm{rep2}}$ is firmly locked by the RF injection locking to improve the comb coherence. The dual-comb signal located in the GHz frequency range is obtained from the multi-heterodyne beating between the two comb lines. To perform the self-reference measurement, the dual-comb signal has to be down-converted to the MHz frequency range due to the restraints of the bandpass filter which is used to filter out one tooth of the dual-comb lines. As shown in the middle panel of Figure \ref{setup}b, the upper curve is the down-converted dual-comb spectrum in the MHz range, while the lower curve is the tooth that is selected by the bandpass filter. Then, the self-reference measurement enables the mixing between the selected tooth and the entire down-converted dual-comb signal, which results in a cancellation of the dual-comb $f_{\rm{ceo}}$ and stability improvement of the dual-comb signal. As shown in the right panel of Figure \ref{setup}b, the obtained self-referenced dual-comb signal consists of lines starting from the line at $f_{\rm{rep}}$ followed by lines at 2$f_{\rm{rep}}$, 3$f_{\rm{rep}}$, 4$f_{\rm{rep}}$, and so on. Due to removal of the dual-comb carrier noise, the whole self-referenced dual-comb signal shows improved stability and coherence, while the intrinsic spectral information is kept.

\section{Results}

\subsection{Frequency comb and dual-comb performance}

The two laser combs used in this work are 6-mm-long THz QCLs with a single plasmon waveguide geometry. The light-current-voltage characteristics of the two lasers are shown in Supplementary Figure S1. At a heat sink temperature of 30 K, both lasers are able to work in continuous wave (cw) mode. The inter-mode beatnote measurement is an efficient and indirect technique to characterize the comb behaviour of QCLs. Benefiting from the ultrafast carrier relaxation in a QCL, the inter-mode beatnote signal can be measured electrically using the QCL itself as a detector. The inter-mode beatnote maps of the two QCL combs with different drive currents are reported in Supplementary figure S2. It can be seen that when the drive current is higher than 850 mA for Comb1 or 830 mA for Comb2, at most currents in the dynamic range, the frequency comb operation can be obtained. For the comb and dual-comb measurements, we operate both lasers at a drive current of 920 mA and the cw output power for each laser is around 1 mW. 

\begin{figure}[!t]
 \centering
 \includegraphics[width=0.9\linewidth]{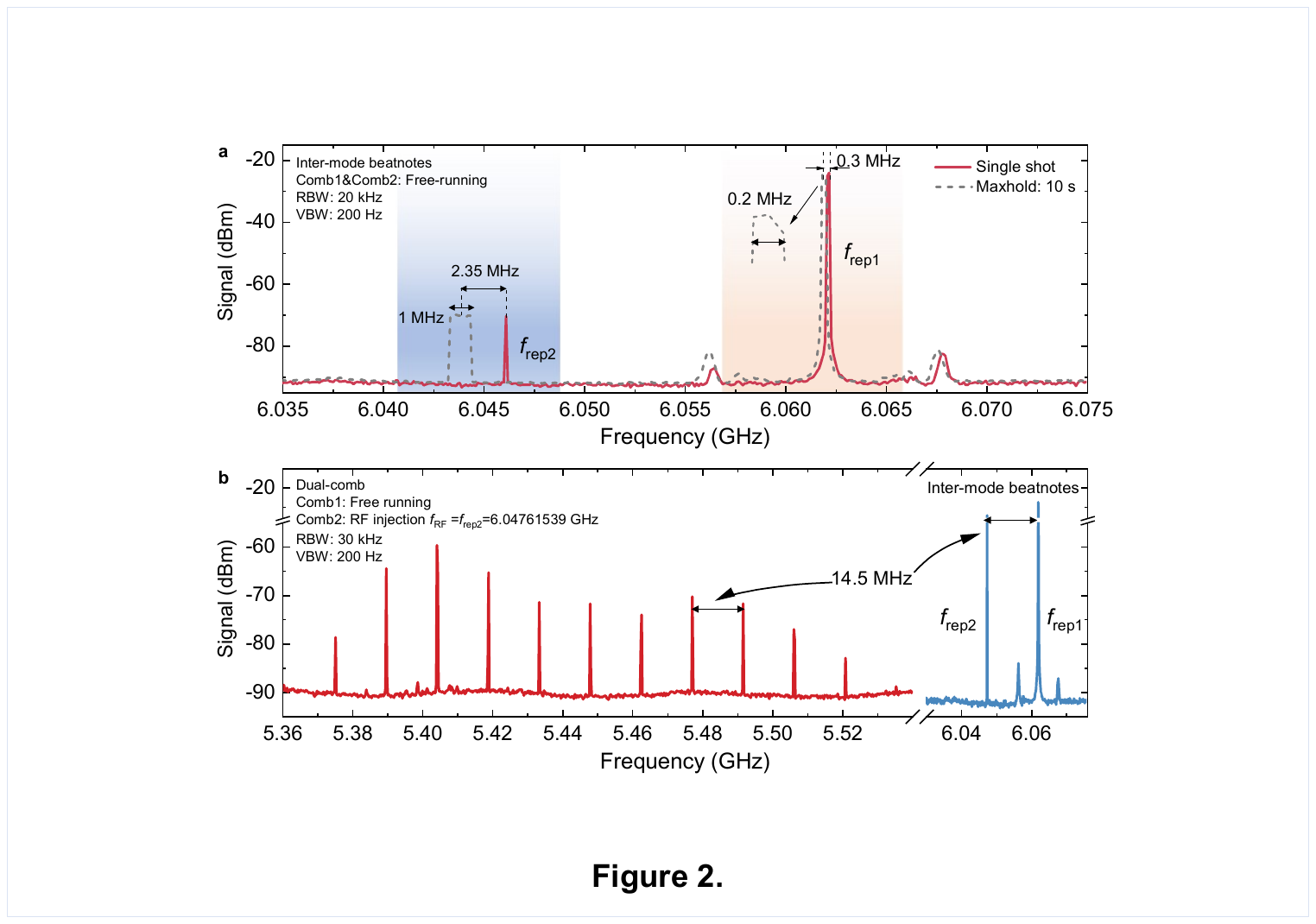}
 \caption{a) Measured inter-mode beatnotes of Comb1 (yellow background) and Comb2 (blue background). The solid and dashed lines are the single shot and ``maxhold" (10 s time duration) traces, respectively. The RBW and VBW parameters are set to 20 kHz and 200 Hz, respectively. b) Dual-comb spectrum (red curve) with the corresponding inter-mode beatnotes (blue curve) measured from the dual-comb source. Comb1 is operated in free-running and Comb2 is locked by the RF injection locking at a frequency of 6.04761539 GHz. The line spacing of the dual-comb signal is equal to the frequency difference of the inter-mode beatnotes. For all measurements, the temperature is stabilized at 29.5 K with a temperature deviation smaller than 50 mK and the drive current is set to 920 mA for both laser combs.}
 \label{comb character}
\end{figure}

In \textbf{Figure \ref{comb character}}a, we show the measured inter-mode beatnotes of Comb1 and Comb2 in single shot (black solid curve) and ``maxhold" modes (grey dashed curve). From the ``maxhold" measurement, we can see that Comb2 shows a higher instability of $f_{\rm{rep}}$: during a time duration of 10 s, the frequency drift is measured to be 1 MHz, while Comb1 shows a drift of 0.2 MHz. When we compare the ``maxhold" spectrum with the single shot one, we can further see that Comb1 is much more stable than Comb2, which is indicated by the measured frequency difference between the center frequency of the ``maxhold" peak and the single shot frequency, i.e., 2.35 MHz for Comb2 and 0.3 MHz for Comb1. It is worth noting that although the nominal design and dimensions for both lasers are identical, the two laser combs show much different comb stability in free-running because all conditions for the two lasers cannot be exactly same. In view of this, to obtain an acceptable dual-comb signal for further self-reference measurements, we implement the microwave injection on Comb2 to stabilize its repetition frequency as shown in Figure \ref{setup}a.

Figure \ref{comb character}b shows the measured multiheterodyne dual-comb spectrum when Comb1 is operated in free-running and Comb2 is stabilized by a microwave injection at a round-trip frequency of 6.04761539 GHz with a RF power of -4 dBm. Note that here a relatively low RF power is used to avoid high phase noise introduced by the RF synthesizer\cite{2019chip}. Equally-spaced dual-comb lines with a spacing of 14.5 MHz can be clearly observed in figure \ref{comb character}b and the dual-comb spacing is equal to the difference between the two comb repetition frequencies, $f_{\rm{rep1}}$ and $f_{\rm{rep2}}$.

\subsection{Self-referenced dual-comb}

After the dual-comb signal is obtained, we can further implement the self-reference method to improve the dual-comb stability. As we explained previously, we select one dual-comb line and mix it with the whole original dual-comb lines to cancel the common carrier frequency noise. To explain the self-reference process in detail, we start from the THz frequencies. If the $N$-th order of modes of the two QCL combs are considered for the generation of the corresponding dual-comb line, the frequencies of the two THz modes from Comb1 and Comb2, respectively, can be written as,
\begin{equation}
  \begin{minipage}[c]{0.9\linewidth}
    \raggedright
    $f_1=f_{\rm{ceo1}}+Nf_{\rm{rep1}}$;
    \par
    $f_2=f_{\rm{ceo2}}+Nf_{\rm{rep2}}$,
  \end{minipage}
  \label{fc}
\end{equation}
where \emph{f}$_{\rm{ceo1}}$ and \emph{f}$_{\rm{ceo2}}$ are the carrier offset frequencies of Comb1 and Comb2, respectively, and $N$ is an integer. Hence, the frequency of the dual-comb line that is resulted from the beating of $f_1$ and $f_2$ can be written as,
\begin{equation}
  \begin{minipage}[c]{0.9\linewidth}
        \raggedright
$f=f_1-f_2=f_{\rm{ceo1}}-f_{\rm{ceo2}}+N(f_{\rm{rep1}}-f_{\rm{rep2}})$.
  \end{minipage}
  \label{dc-1}
\end{equation}
Alternatively, we can rewrite it in the similar form as optical frequency combs,
\begin{equation}
  \begin{minipage}[c]{0.9\linewidth}
        \raggedright
         $f=f_{\rm{ceo,dc}}+Nf_{\rm{rep,dc}}$,
  \end{minipage}
  \label{dc-2}
\end{equation}
where $f_{\rm{ceo,dc}}$=$f_{\rm{ceo1}}$-$f_{\rm{ceo2}}$ and $f_{\rm{rep,dc}}$=$f_{\rm{rep1}}$-$f_{\rm{rep2}}$ are the carrier offset frequency and repetition frequency of a dual-comb signal, respectively. From equations \ref{dc-1} and \ref{dc-2}, we can say that the noise of a dual-comb signal is from its carrier offset frequency and repetition frequency noises, which is actually equivalent to the carrier offset noise and repetition frequency noise of Comb1 against those of Comb2. Considering jitters for carrier offset and repetition frequencies, noise terms can be added to the steady terms as follows,
\begin{equation}
  \begin{minipage}[c]{0.9\linewidth}
        \raggedright
$f=f_{\rm{ceo,dc}}+\delta f_{\rm{ceo,dc}}+N(f_{\rm{rep,dc}}+\delta f_{\rm{rep,dc}})$,
  \end{minipage}
  \label{dc-jitter}
\end{equation}

where $\delta f_{\rm{ceo,dc}}$ and $\delta f_{\rm{rep,dc}}$ are the noise terms (fluctuations) of $f_{\rm{ceo,dc}}$ and $f_{\rm{rep,dc}}$, respectively. The frequency of each dual-comb line can be obtained from equation \ref{dc-jitter} by varying the integer $N$. Note that all dual-comb lines share the common noise terms of $\delta f_{\rm{ceo,dc}}$ and accumulated $\delta f_{\rm{rep,dc}}$. To implement the self-reference measurement, one dual-comb line is chosen, for example, the $M$-th order line ($f_M$), which can be written as,
\begin{equation}
  \begin{minipage}[c]{0.9\linewidth}
        \raggedright
        $f_M=f_{\rm{ceo,dc}}+\delta f_{\rm{ceo,dc}}+M(f_{\rm{rep,dc}}+\delta f_{\rm{rep,dc}})$.
  \end{minipage}
  \label{dc-line}
\end{equation}
Consequently, the beating between $f_M$ and each $f$ results in the self-referenced dual-comb signal. The newly generated frequency ($f_{\rm{sr}}$) can be expressed,
\begin{equation}
  \begin{minipage}[c]{0.9\linewidth}
        \raggedright
$f_{\rm{sr}}=f-f_M=(N-M)(f_{\rm{rep,dc}}+\delta f_{\rm{rep,dc}})=(N-M)f_{\rm{rep,dc}}+(N-M)\delta f_{\rm{rep,dc}}$.
  \end{minipage}
  \label{dc-self}
\end{equation}
Note that in equation \ref{dc-self}, $M$ is a constant integer and $N$ is a variable integer. From equation \ref{dc-self}, we can find that the common noise of $\delta f_{\rm{ceo,dc}}$ is completely eliminated during the self-mixing process. Furthermore, the coefficient of the noise term of $\delta f_{\rm{rep,dc}}$ is also reduced from $N$ to $N-M$ after the self-reference process. It has to be clarified that for a frequency comb, $f_{\rm{ceo}}$ is normally small (close to DC) and $N$ is very big. Therefore, in free running, the noise of different comb lines are comparable without sizable differences. However, once the self-reference is implemented, we can see that the noise terms of $f_{\rm{ceo,dc}}$ and $f_{\rm{rep,dc}}$ are either removed or significantly suppressed.

\begin{figure}[!t]
 \centering
 \includegraphics[width=0.9\linewidth]{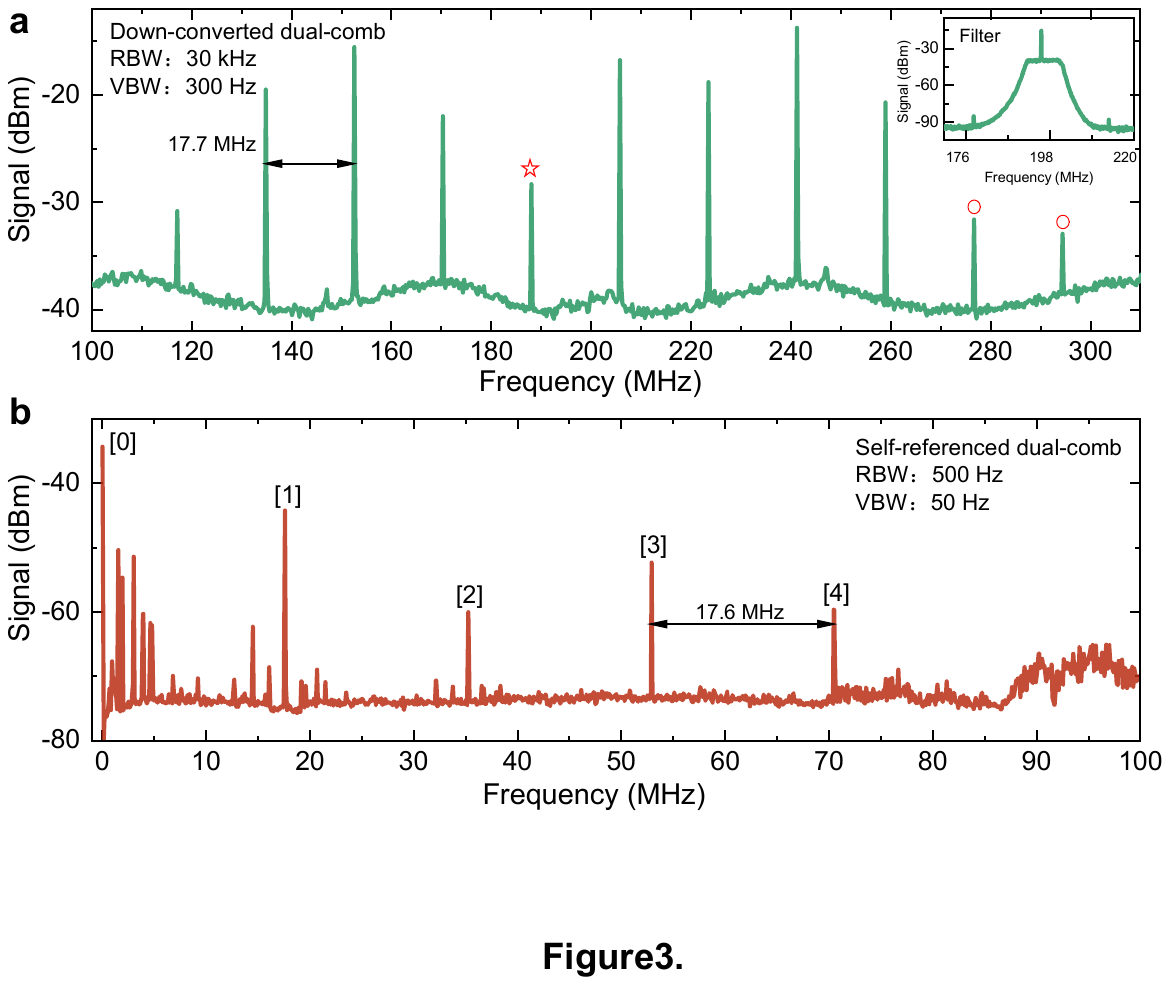}
 \caption{Down-converted and self-referenced dual-comb spectra. a) Down-converted dual-comb spectrum measured with a RBW of 30 kHz and a VBW of 300 Hz. The dual-comb line spacing is 17.7 MHz. The tooth that is selected by a bandpass filter for the self-reference mixing is marked by a red star. Two lines at high frequencies marked by red circles with a lower power level don't participate in the self-reference mixing with the line marked by the red star. The inset shows the typical spectrum of the filtered dual-comb tooth measured with the same RBW and VBW parameters. b) Self-referenced dual-comb spectrum measured with a RBW of 500 Hz and a VBW of 50 Hz. 5 lines marked by [0]-[4] are equally spaced by 17.6 MHz which is consistent with the line spacing shown in \textbf{a}.}
 \label{self-reference}
\end{figure}

The experimental results of the self-referenced dual-comb are shown in \textbf{Figure \ref{self-reference}}. To perform the self-reference mixing, we have to first down-convert the dual-comb signal as shown in figure \ref{comb character}b to lower frequencies around 200 MHz. This frequency down-conversion is necessary because the bandpass filter that is used to filter out one of the dual-comb lines is only working at lower frequencies. A RF signal at 5.35176 GHz with a power of 7 dBm is used to mix with the dual-comb signal in GHz range and bring it down to MHz range. Figure \ref{self-reference}a shows the down-converted dual-comb spectrum with 11 lines and the line spacing is 17.7 MHz measured with a resolution bandwidth (RBW) of 30 kHz and a video bandwidth (VBW) of 300 Hz. The inset of figure \ref{self-reference}a is the spectrum of the selected line at 195.7 MHz by using a tunable bandpass filter with a bandwidth of 8 MHz (around 5\% of the selected center frequency). It can be seen that the frequency of the selected line doesn't perfectly match with any line shown in figure \ref{self-reference}a. This is because that the two spectra were taken at different times and the addition of a bandpass filter in the electrical circuit can change the absolute frequency positions of the lines. However, the frequency mismatch between the two spectra is not a problem for the self-reference measurement because for the self-reference mixing the instantaneous frequency of the selected line can always find its original one from the down-converted dual-comb spectrum and at each time the two frequencies are exactly same. Therefore, the proposed self-reference measurement is featured by the immunity of long-term noises in the dual-comb system. 

Figure \ref{self-reference}b shows the self-referenced dual-comb spectrum measured with a RBW of 500 Hz and a VBW of 50 Hz. Since the selected line (inset of figure \ref{self-reference}a) is from the central region of the dual-comb signal (indicated by a red star in figure \ref{self-reference}a), in the self-referenced dual-comb spectrum we only observe 5 lines which are equally spaced by 17.6 MHz. Here, the 5 lines actually correspond to 9 dual-comb lines in figure \ref{self-reference}a: the first line at 0 Hz corresponds to beating of the selected line and itself in the dual-comb signal ($N$=$M$, see equation \ref{dc-self}); the other lines in figure \ref{self-reference}b are resulted from two beatings of the selected line with its right and left neighbouring lines in the dual-comb spectra ($N{\neq}M$, see equation \ref{dc-self}). Note that the self-referenced signal doesn't include the information of the two dual-comb lines with highest frequencies (marked by circles in figure \ref{self-reference}b), probably due to the higher noise as the frequency is greater than 86 MHz or the low power level of the two lines. One can also find that the measured line spacing of the self-referenced dual-comb (figure \ref{self-reference}b) is slightly different from the down-converted dual-comb (figure \ref{self-reference}a), which is due to the instability of $f_{\rm{rep,dc}}$ originally from the instability of $f_{\rm{rep}}$ of the two QCL combs. It is worth mentioning that in figure \ref{self-reference}b in the vicinity of lines marked by [0], [1], and [2], several redundant lines can be clearly observed. We measured the line spacing of these lines and found that they were not equally separated. Therefore, we assume the low frequency disordered lines are generated by the microwave amplifiers in the electrical circuit of the measurement system rather than from the laser combs because the disordered lines are absent in Figs. \ref{comb character}b and \ref{self-reference}a.

\begin{figure}[!t]
 \centering
 \includegraphics[width=0.85\linewidth]{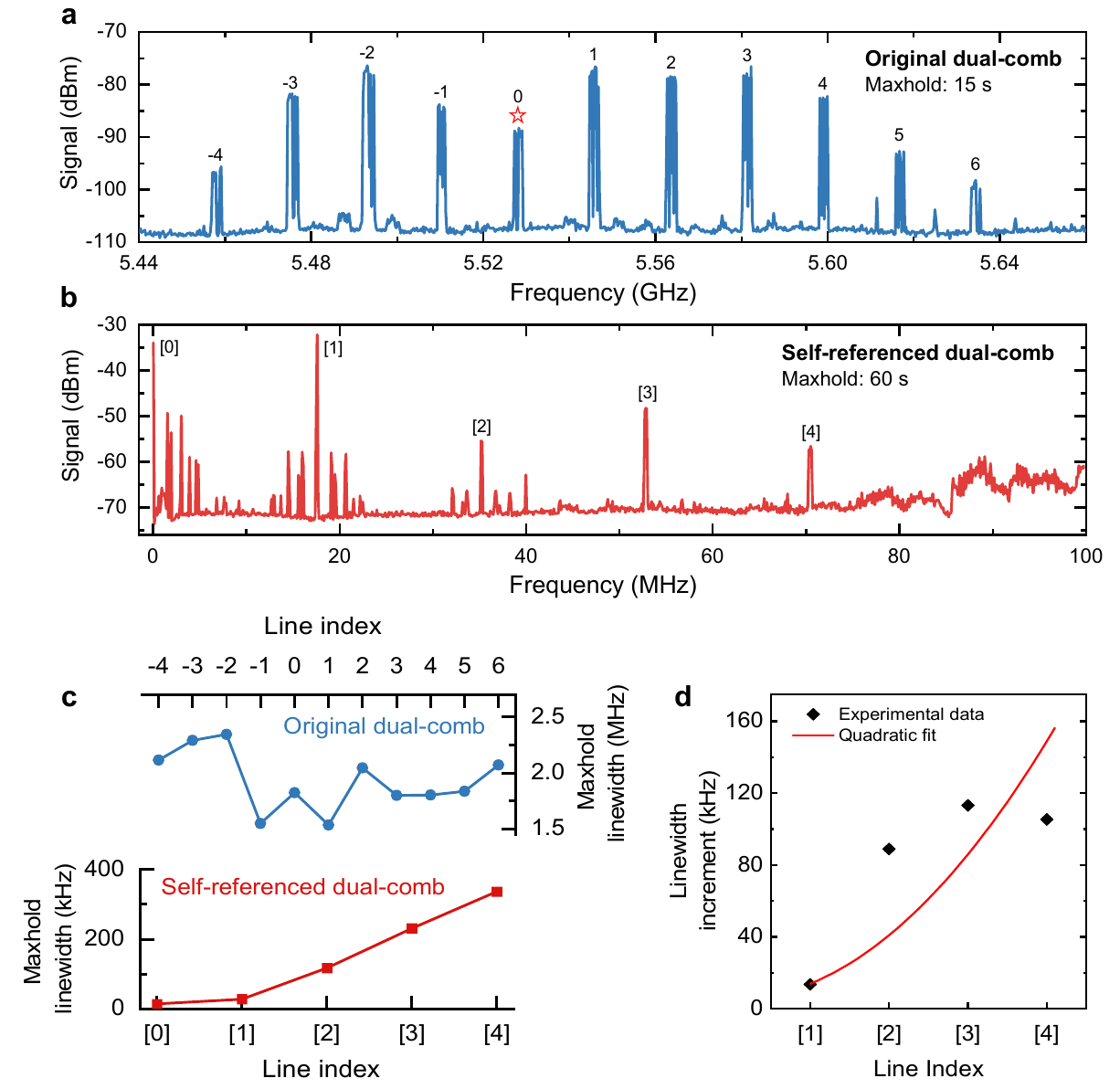}
  \caption{``Maxhold" measurement for the dual-comb spectra with and without the self-reference technique. a) Original dual-comb spectrum without the self-reference method measured with a ``maxhold“ duration time of 15 seconds. The line with an index '0' marked by a red star is the selected tooth for the self-reference mixing. b) Self-referenced dual-comb spectrum measured a ``maxhold" duration time of 60 seconds. For both measurements, RBW and VBW parameters are set to 500 Hz and 50 Hz, respectively. c) ``Maxhold" linewidths of the original (blue) and self-referenced (red) dual-comb lines extracted from a) and b). d) Linewidth increment as a function of line index for the self-referenced dual-comb lines extracted from c). The diamonds show the measured results and the red curve is a quadratic fit. }
 \label{maxhold}
\end{figure}

In figure \ref{self-reference}, we show that the self-referenced dual-comb signal can be obtained as expected by following the experimental setup shown in figure \ref{setup}. To further demonstrate how stable the self-referenced dual-comb signal is, we perform the ``maxhold" measurement for the dual-comb signals with and without the self-reference technique. The main results are shown in \textbf{figure \ref{maxhold}}. Here, when the ``maxhold" function of the spectrum analyser is switched on, the maxima of the spectral peaks can be stored during the measurement. Therefore, the ``maxhold" spectra can visually show the long-term stability of a signal. Figure \ref{maxhold}a shows the ``maxhold" dual-comb spectrum without the self-reference in GHz range for a time duration of 15 s. It can be seen that for all 11 lines, the frequency drifts during 15 s are comparable and lie in a level of 2 MHz (see figure \ref{maxhold}c) which show a good agreement with the ``maxhold" linewidths measured from another QCL dual-comb system with a laser cavity of 5.5 mm\cite{zhao2021active}. 

In figure \ref{maxhold}b, we show the ``maxhold" trace of the self-referenced dual-comb signal measured for a time duration of 60 s. For the self-reference process, the line with an index of 0 marked by a red star in figure \ref{maxhold}a is selected to mix with the entire dual-comb signal. A significant stability improvement can be clearly observed from figure \ref{maxhold}b: for 60 s that is four times longer than the time duration used for the original dual-comb measurement, a ``maxhold" linewidth narrowing for all lines marked by [0], [1], [2], [3], and [4] can be seen by the naked eyes. The measured linewidths for both traces in Figs. \ref{maxhold}a and \ref{maxhold}b are summarized in figure \ref{maxhold}c. One can see a monotonous increase of the ``maxhold" linewidth with line index for the self-referenced dual-comb, while the original dual-comb shows similar linewidths around 2 MHz. Note that the narrowest ``maxhold" linewidths for line [0] and [1] are 14.8 kHz and 28.5 kHz, respectively. As we explained, the reason for the significant reduction of the linewidth of the self-referenced dual-comb signal is the cancellation of the carrier frequency and reduced repetition frequency noise of the dual-comb signal. To further show the ``maxhold" linewidth comparison between the self-referenced dual-comb and original dual-comb spectra, more experimental data were measured with different time durations as shown in Supplementary Figs. S3 and S4. To clearly show the dependence of the ``maxhold" linewidth on the line index, we plot the linewidth increment as a function of line index in figure \ref{maxhold}d. The diamonds are measured results and the red line shows a quadratic fit\cite{zhao2021active} to the experimental data. For the given 4 experimental points, we can see the measured linewidth increments fairly follow a quadratic dependence on line index.

\section{Discussion}

The results shown in Figs. \ref{self-reference} and \ref{maxhold} experimentally prove that the self-reference method can be successfully implemented onto a QCL dual-comb system to significantly improve the dual-comb stability. It has to be noted that the proposed self-reference technique doesn't use traditional locking hardwares to control the carrier offset and repetition frequencies of the two QCLs. Therefore, the laser combs are still running with large noises, see Figs. \ref{comb character}a and \ref{maxhold}a. Although it is not a physical locking technique, the self-reference method can largely improve the dual-comb signal stability due to its noise cancellation ability as we elaborated in the main text. This is important for reliably reading the experimental data for some spectroscopic applications, especially when the absorption linewidth of a sample is wider than the long-term frequency drift of the THz modes (which is around 2 MHz as shown in Figs. \ref{maxhold}a and \ref{maxhold}c). Generally, this can be satisfied for most of cases. Furthermore, the self-reference method is able to maximally simplify the experimental setup for compact applications.  

In the current self-reference experiment, to obtain a higher signal to noise ratio, we select a dual-comb line with a high power located in the central range of the spectra for self-mixing. Therefore, each self-referenced dual-comb line except the line marked by an index [0] are resulted from two different beating processes. So, the mode aliasing exists in the self-referenced dual-comb spectrum. This is the reason that the number of the self-referenced dual-comb lines is less than that of the original dual-comb lines (see Figs. \ref{self-reference} and \ref{maxhold}). In this situation, the one-to-one correspondence cannot be obtained between the referenced dual-comb and original dual-comb spectra. In this work, the main objective is to demonstrate the operation principle and feasibility of the self-reference method rather than spectroscopic applications. For a practical spectroscopic application, the mode aliasing can be avoided with more effort. For example, we can further amplify the original dual-comb signal and select the leftmost or rightmost line to perform the self-mixing experiment. Then, the self-referenced dual-comb spectra with a precise one-to-one correspondence can be obtained.  

From figure \ref{maxhold}, we can see that by implementing the self-reference technique the dual-comb signal is significantly improved. However, the dual-comb repetition frequency noise is still present and its quadratic increase with the line index can be clearly observed. Actually, the dual-comb repetition frequency noise can be further suppressed. For example, a microwave double-injection\cite{2019chip} can be applied onto the two laser combs to lock the two repetition frequencies simultaneously. In this case, the repetition frequency noises of the two laser combs are completely removed. Consequently, the dual-comb repetition frequency noise is accordingly removed. 

Finally, it is worth mentioning that the proposed self-reference method is currently implemented onto a THz QCL dual-comb source. However, the general method can be also adopted for other dual-comb sources or even for a single QCL frequency comb. In principle, for a single THz QCL comb, if a narrow band filter and a broadband mixer working in the terahertz range are available, the self-mixing between the selected comb line and the entire frequency comb signal can significantly improve the long-term stability of the frequency comb.

\section{Conclusion}

In summary, we have demonstrated, at the first time as far as we know, a self-reference method for a pure QCL dual-comb system working around 4.2 THz. Without physical locking hardwares, the proposed self-reference technique can significantly improve the dual-comb signal stability by eliminating the dual-comb carrier offset frequency and reducing the dual-comb repetition frequency noise. Experimental results revealed that all self-referenced dual-comb lines showed much improved long-term stability than the original dual-comb lines. The measured narrowest ``maxhold" linewidth of the self-referenced signal is 14.8 kHz (60 s time duration), which is 130 times smaller than that measured from the original dual-comb signal (15 s time duration). The simplified self-reference technique can be further employed for the fast THz dual-comb spectroscopy especially when the absorption linewidth of a sample is wider than the long-term frequency drift of the THz modes. In principle, the self-reference method can be also implemented in other laser systems to improve the system stability for various applications, such as, spectroscopy, imaging, communications, etc.


\section{Experimental Section}
\threesubsection{THz QCLs and characterizations}
The THz QCLs used in this work is based on a GaAs/AlGaAs hybrid active region structure in which the bound-to-continuum transitions are employed for the THz photon emission and resonant-phonon scatterings are designed for the depopulation of the lower laser state. The detailed layer thickness and doping profiles can be found in ref.\cite{WanSR}. The entire active region of the QCL was grown on a semi-insulating GaAs(100) substrate by a molecular beam epitaxy (MBE) system. Then, the grown wafer was processed into single plasmon waveguide laser ridges with a ridge width of 150 $\mu$m. Finally, 6-mm-long laser ridges were cleaved and indium-bonded onto copper heat sinks for the wire bonding.

For electrical and optical characterizations, the QCL devices were mounted onto a cold finger of a cryostat. The average power of the lasers was measured using a power meter (Ophir, 3A-P THz) with a pair of parabolic mirrors for light collection and collimation. The inter-mode beatnote of the QCL comb, as well as the multiheterodyne dual-comb signals, was measured using the device itself as a detector and the microwave spectra can be directly registered on a spectrum analyser (Rohde \& Schwarz, FSW26). \\
\threesubsection{Multiheterodyne self-detection}
In the work, the main THz detection mechanism is the multiheterodyne self-detection. As shown in figure \ref{setup}, when the light of Comb2 is incident to Comb1, the frequency of Comb1 will be modulated by Comb2. In our experimental setup, the modulation of frequency multiheterodyne will be visually shown by the modulation of the drive current of Comb1. The ultra-fast carrier relaxation time in picoseconds level of THz QCLs gives them the ability to act as a fast THz detector\cite{LiOE} for multiheterodyne measurements. Therefore, by probing the current of Comb1, we are able to extract the multiheterodyne information of the two lasers. Note that, the two lasers, Comb1 and Comb2, have the equivalent function of light emission and detection. Therefore, either laser can be used as a fast detector\cite{2019chip,LiACSPhoton}. In this work, Comb1 is operated as a fast detector for the multiheterodyne detection. The AC port of a Bias-T (Marki, BT2-0026) that is directly connected to Comb1 can extract the high frequency inter-mode beatnote and dual-comb signals.\\

\threesubsection{Self-reference} 
As explained in the main paper, the proposed self-reference method is implemented by selecting one of the dual-comb lines and beating it with the entire dual-comb lines to eliminate the common carrier offset frequencies and reducing the dual-comb repetition frequency noise. Without using any external locking facilities, the self-reference technique can significantly improve the observed dual-comb signal stability. As shown in figure \ref{setup}, the original dual-comb signal in the GHz range is first measured via the AC port of a Bias-T (Marki, BT2-0026) that is connected to Comb1 (detector). Before it is sent to a mixer (Mini-Circuit, ZX05-73L-S+) for the frequency down-conversion to the MHz range, the dual-comb signal is amplified using a microwave amplifier (CONNPHY, CLN-1G18G-4030-S) with a gain of 40 dB. Then, the down-converted dual-comb signal in the MHz range is split into two using a microwave coupler for the self-reference mixing. One of the splitted signals is sent to a tunable bandpass filter (K{\&}L Microwave, 5BT-95/190-5-N/N) for the one line selection. Before the self-reference frequency mixing, the two signals are amplified again. Finally, the filtered signal (LO) and entire dual-comb signal (RF) are mixed in a microwave mixer (Mini-Circuit, ZX05-1L-S+) which then generates the carrier-frequency-free dual-comb spectrum with a significant noise suppression. \\

\medskip
\textbf{Supporting Information} \par 
Supporting Information is available from the author.

\medskip
\textbf{Acknowledgements} \par 
Z.L. and X.M. contributed equally to this work. This work is supported by the National Natural Science Foundation of China (61875220, 61927813, 62035005, 61991430, 62105351), the ``From 0 to 1" Innovation Program of the Chinese Academy of Sciences (ZDBS-LY-JSC009), the Scientific Instrument and Equipment Development Project of the Chinese Academy of Sciences (Grant No. YJKYYQ20200032), the National Science Fund for Excellent Young Scholars (62022084), Shanghai Outstanding Academic Leaders Plan (20XD1424700), and Shanghai Youth Top Talent Support Program.

\medskip

%
\bibliographystyle{MSP}
\bibliography{REF.bib}

\begin{thebibliography}{10}
\providecommand{\url}[1]{\texttt{#1}}
\providecommand{\urlprefix}{URL }

\bibitem{bernhardt2010cavity}
B.~Bernhardt, A.~Ozawa, P.~Jacquet, M.~Jacquey, Y.~Kobayashi, T.~Udem,
  R.~Holzwarth, G.~Guelachvili, T.~W. H{\"a}nsch, N.~Picqu{\'e},
\newblock \emph{Nature Photonics} \textbf{2010}, \emph{4}, 1 55.

\bibitem{villares2014dual}
G.~Villares, A.~Hugi, S.~Blaser, J.~Faist,
\newblock \emph{Nature Communications} \textbf{2014}, \emph{5}, 1 5192.

\bibitem{CoddingtonOptica}
I.~Coddington, N.~Newbury, W.~Swann,
\newblock \emph{Optica} \textbf{2016}, \emph{3}, 4 414.

\bibitem{dutt2018chip}
A.~Dutt, C.~Joshi, X.~Ji, J.~Cardenas, Y.~Okawachi, K.~Luke, A.~L. Gaeta,
  M.~Lipson,
\newblock \emph{Science Advances} \textbf{2018}, \emph{4}, 3 e1701858.

\bibitem{picque2019frequency}
N.~Picqu{\'e}, T.~W. H{\"a}nsch,
\newblock \emph{Nature Photonics} \textbf{2019}, \emph{13}, 3 146.

\bibitem{2016terahertz}
Y.~Yang, D.~Burghoff, D.~J. Hayton, J.-R. Gao, J.~L. Reno, Q.~Hu,
\newblock \emph{Optica} \textbf{2016}, \emph{3}, 5 499.

\bibitem{ideguchi2014adaptive}
T.~Ideguchi, A.~Poisson, G.~Guelachvili, N.~Picqu{\'e}, T.~W. H{\"a}nsch,
\newblock \emph{Nature Communications} \textbf{2014}, \emph{5}, 1 3375.

\bibitem{millot2016frequency}
G.~Millot, S.~Pitois, M.~Yan, T.~Hovhannisyan, A.~Bendahmane, T.~W. H{\"a}nsch,
  N.~Picqu{\'e},
\newblock \emph{Nature Photonics} \textbf{2016}, \emph{10}, 1 27.

\bibitem{zhao2016picometer}
X.~Zhao, G.~Hu, B.~Zhao, C.~Li, Y.~Pan, Y.~Liu, T.~Yasui, Z.~Zheng,
\newblock \emph{Optics Express} \textbf{2016}, \emph{24}, 19 21833.

\bibitem{LiACSPhoton}
H.~Li, Z.~Li, W.~Wan, K.~Zhou, X.~Liao, S.~Yang, C.~Wang, J.~C. Cao, H.~Zeng,
\newblock \emph{ACS Photonics} \textbf{2020}, \emph{7}, 1 49.

\bibitem{link2017dual}
S.~M. Link, D.~Maas, D.~Waldburger, U.~Keller,
\newblock \emph{Science} \textbf{2017}, \emph{356}, 6343 1164.

\bibitem{ideguchi2012adaptive}
T.~Ideguchi, A.~Poisson, G.~Guelachvili, T.~W. H{\"a}nsch, N.~Picqu{\'e},
\newblock \emph{Optics Letters} \textbf{2012}, \emph{37}, 23 4847.

\bibitem{yan2017mid}
M.~Yan, P.-L. Luo, K.~Iwakuni, G.~Millot, T.~W. H{\"a}nsch, N.~Picqu{\'e},
\newblock \emph{Light: Science \& Applications} \textbf{2017}, \emph{6}, 10
  e17076.

\bibitem{kohler2002terahertz}
R.~K{\"o}hler, A.~Tredicucci, F.~Beltram, H.~E. Beere, E.~H. Linfield, A.~G.
  Davies, D.~A. Ritchie, R.~C. Iotti, F.~Rossi,
\newblock \emph{Nature} \textbf{2002}, \emph{417}, 6885 156.

\bibitem{burghoff2014terahertz}
D.~Burghoff, T.-Y. Kao, N.~Han, C.~W.~I. Chan, X.~Cai, Y.~Yang, D.~J. Hayton,
  J.-R. Gao, J.~L. Reno, Q.~Hu,
\newblock \emph{Nature Photonics} \textbf{2014}, \emph{8}, 6 462.

\bibitem{hugi2012mid}
A.~Hugi, G.~Villares, S.~Blaser, H.~Liu, J.~Faist,
\newblock \emph{Nature} \textbf{2012}, \emph{492}, 7428 229.

\bibitem{2017multiwatt}
L.~H. Li, L.~Chen, J.~R. Freeman, M.~Salih, P.~Dean, A.~G. Davies, E.~H.
  Linfield,
\newblock \emph{Electronics Letters} \textbf{2017}, \emph{53}, 12 799.

\bibitem{2018heterogeneous}
M.~R{\"o}sch, M.~Beck, M.~J. S{\"u}ess, D.~Bachmann, K.~Unterrainer, J.~Faist,
  G.~Scalari,
\newblock \emph{Nanophotonics} \textbf{2018}, \emph{7}, 1 237.

\bibitem{2012ground}
C.~W.~I. Chan, Q.~Hu, J.~L. Reno,
\newblock \emph{Applied Physics Letters} \textbf{2012}, \emph{101}, 15 151108.

\bibitem{2016chip}
M.~R{\"o}sch, G.~Scalari, G.~Villares, L.~Bosco, M.~Beck, J.~Faist,
\newblock \emph{Applied Physics Letters} \textbf{2016}, \emph{108}, 17 171104.

\bibitem{2019chip}
Z.~Li, W.~Wan, K.~Zhou, X.~Liao, S.~Yang, Z.~Fu, J.~Cao, H.~Li,
\newblock \emph{Physical Review Applied} \textbf{2019}, \emph{12}, 4 044068.

\bibitem{villares2015chip}
G.~Villares, J.~Wolf, D.~Kazakov, M.~J. S{\"u}ess, A.~Hugi, M.~Beck, J.~Faist,
\newblock \emph{Applied Physics Letters} \textbf{2015}, \emph{107}, 25 251104.

\bibitem{zhou2019ridge}
K.~Zhou, H.~Li, W.~Wan, Z.~Li, X.~Liao, J.~Cao,
\newblock \emph{Applied Physics Letters} \textbf{2019}, \emph{114}, 19 191106.

\bibitem{villares2016dispersion}
G.~Villares, S.~Riedi, J.~Wolf, D.~Kazakov, M.~J. S{\"u}ess, P.~Jouy, M.~Beck,
  J.~Faist,
\newblock \emph{Optica} \textbf{2016}, \emph{3}, 3 252.

\bibitem{barbieri200713}
S.~Barbieri, W.~Maineult, S.~S. Dhillon, C.~Sirtori, J.~Alton, N.~Breuil, H.~E.
  Beere, D.~A. Ritchie,
\newblock \emph{Applied Physics Letters} \textbf{2007}, \emph{91}, 14 143510.

\bibitem{2010injection}
P.~Gellie, S.~Barbieri, J.~F. Lampin, P.~Filloux, C.~Manquest, C.~Sirtori,
  I.~Sagnes, S.~P. Khanna, E.~H. Linfield, A.~G. Davies, et~al.,
\newblock \emph{Optics Express} \textbf{2010}, \emph{18}, 20 20799.

\bibitem{li2019graphene}
H.~Li, M.~Yan, W.~Wan, T.~Zhou, K.~Zhou, Z.~Li, J.~Cao, Q.~Yu, K.~Zhang, M.~Li,
  et~al.,
\newblock \emph{Advanced Science} \textbf{2019}, \emph{6}, 20 1900460.

\bibitem{holzwarth2000optical}
R.~Holzwarth, T.~Udem, T.~W. H{\"a}nsch, J.~Knight, W.~Wadsworth, P.~S.~J.
  Russell,
\newblock \emph{Physical Review Letters} \textbf{2000}, \emph{85}, 11 2264.

\bibitem{barbieri2010phase}
S.~Barbieri, P.~Gellie, G.~Santarelli, L.~Ding, W.~Maineult, C.~Sirtori,
  R.~Colombelli, H.~Beere, D.~Ritchie,
\newblock \emph{Nature Photonics} \textbf{2010}, \emph{4}, 9 636.

\bibitem{oustinov2010phase}
D.~Oustinov, N.~Jukam, R.~Rungsawang, J.~Mad{\'e}o, S.~Barbieri, P.~Filloux,
  C.~Sirtori, X.~Marcadet, J.~Tignon, S.~Dhillon,
\newblock \emph{Nature Communications} \textbf{2010}, \emph{1}, 1 69.

\bibitem{zhao2021active}
Y.~Zhao, Z.~Li, K.~Zhou, X.~Liao, W.~Guan, W.~Wan, S.~Yang, J.~Cao, D.~Xu,
  S.~Barbieri, et~al.,
\newblock \emph{Laser \& Photonics Reviews} \textbf{2021}, \emph{15}, 4
  2000498.

\bibitem{burghoff2016computational}
D.~Burghoff, Y.~Yang, Q.~Hu,
\newblock \emph{Science Advances} \textbf{2016}, \emph{2}, 11 e1601227.

\bibitem{chen2020adaptive}
J.~Chen, K.~Nitta, X.~Zhao, T.~Mizuno, T.~Minamikawa, F.~Hindle, Z.~Zheng,
  T.~Yasui,
\newblock \emph{Advanced Photonics} \textbf{2020}, \emph{2}, 3 036004.

\bibitem{cappelli2019retrieval}
F.~Cappelli, L.~Consolino, G.~Campo, I.~Galli, D.~Mazzotti, A.~Campa,
  M.~Siciliani~de Cumis, P.~Cancio~Pastor, R.~Eramo, M.~R{\"o}sch, et~al.,
\newblock \emph{Nature Photonics} \textbf{2019}, \emph{13}, 8 562.

\bibitem{LiOE}
H.~Li, P.~Laffaille, D.~Gacemi, M.~Apfel, C.~Sirtori, J.~Leonardon,
  G.~Santarelli, M.~Rösch, G.~Scalari, M.~Beck, J.~Faist, W.~Hänsel,
  R.~Holzwarth, S.~Barbieri,
\newblock \emph{Optics Express} \textbf{2015}, \emph{23}, 26 33270.

\bibitem{WanSR}
W.~Wan, H.~Li, T.~Zhou, J.~Cao,
\newblock \emph{Scientific Reports} \textbf{2017}, \emph{7} 44109.

\end{thebibliography}




\end{document}